\documentclass[useAMS,usenatbib]{mn2e}
\usepackage{graphicx,rotate,url,mathptmx}
\def\gtsim{\mathrel{\hbox{\rlap{\hbox{\lower4pt\hbox{$\sim$}}}\hbox{$>$}}}}
\def\lesssim{\mathrel{\hbox{\rlap{\hbox{\lower4pt\hbox{$\sim$}}}\hbox{$<$}}}}
\def\Msunpyr{M$_{\odot}\,$yr$^{-1}$}
\def\Msun{M$_{\odot}$}
\def\cm{{\rm\thinspace cm}}
\def\erg{{\rm\thinspace erg}}

\def\Hz{{\rm\thinspace Hz}}

\def\keV{{\rm\thinspace keV}}
\def\km{{\rm\thinspace km}}

\def\Msun{\hbox{$\rm\thinspace M_{\odot}$}}

\def\s{{\rm\thinspace s}}
\def\ps{{\rm\thinspace s^{-1}}}
\def\W{{\rm\thinspace W}}
\def\yr{{\rm\thinspace yr}}
\def\keVscm{\hbox{$\keV\cm^{2}\,$}}

\def\ergpscmps{\hbox{$\erg\cm^{-2}\s^{-1}\,$}}

\def\ergps{\hbox{$\erg\s^{-1}\,$}}

\def\kmps{\hbox{$\km\ps\,$}}

\def\Msunpyr{\hbox{$\Msun\yr^{-1}\,$}}

\def\pscm{\hbox{$\cm^{-2}\,$}}

\def\pccm{\hbox{$\cm^{-3}\,$}}
\def\pscm{\hbox{$\cm^{-2}\,$}}

\def\WHz{\hbox{$\W\Hz\,$}}
\newcommand*{\satellite}[1]{\textit{#1}}
\newcommand*{\xmm}{\satellite{XMM-Newton}}
\newcommand*{\chandra}{\satellite{Chandra}}
\newcommand*{\asca}{\satellite{ASCA}}
\newcommand*{\rosat}{\satellite{ROSAT}}
\newcommand*{\prog}[1]{\textsc{#1}}
\newcommand*{\mysub}[2]{\ensuremath{#1_{\mathrm{#2}}}}

\title{The Galaxy Cluster Abell~3581 as seen by \chandra{}}

\author[Johnstone, et al.]
       {R.M. Johnstone$^1$\thanks{E-mail: rmj@ast.cam.ac.uk}, 
        A.C. Fabian$^1$,  R.G. Morris$^1$ and G.B. Taylor$^2$\\
        $^1$Institute of Astronomy, University of Cambridge, Madingley Road,
        Cambridge CB3 0HA\\
        $^2$NRAO, Socorro, NM 87801, USA}
\date{
      Received }

\pagerange{\pageref{firstpage}--\pageref{lastpage}}
\pubyear{}

\begin{document}

\maketitle

\label{firstpage}

\begin{abstract}

\noindent
We present results from an analysis of a \chandra{} observation of
the cluster of galaxies Abell~3581. We discover the presence of a
point-source in the central dominant galaxy which is coincident with
the core of the radio source PKS~1404-267. The emission from the
intracluster medium is analysed, both as seen in projection on the sky,
and after correcting for projection effects, to determine the
spatial distribution of gas temperature, density and
metallicity. We find that the cluster, despite hosting a moderately
powerful radio source, shows a temperature decline to around
$0.4~\mysub{T}{max}$ within the central 5~kpc.
The cluster is notable
for the low entropy within its core. We test and validate the
\prog{xspec} \texttt{projct}
model for determining the intrinsic cluster gas properties.

\end{abstract}


\section{Introduction}
\label{intro}

Abell~3581 is a nearby ($z\sim 0.0218$), richness class 0,
cluster of galaxies. The central dominant galaxy
IC~4374 has an extensive optical emission-line filament system
(\citealt{DanzigerFocardi88}, \citealt{JFN87}) like that seen around the
central galaxy in more massive clusters (e.g. the Perseus
cluster, \citealt{Conseliceetal01}).

Abell~3581 was previously studied in
X-rays, using data from the \rosat{} and \asca{} satellites by
\citet{Johnstoneetal98} who found that it had a cool intracluster
medium (ICM) with $kT\sim 2$~keV, and a short radiative cooling time
throughout its core. Those data admitted a classical cooling flow with a
significant amount of intrinsic absorption.

Recently, (e.g. \citealt{Petersonetal01}, \citealt{Johnstoneetal02},
\citealt{Petersonetal03},
\citealt{Kaastraetal04}, \citealt{Sandersetal04}) it has been shown
that clusters which have short central cooling times have much less
gas at temperatures below about a third of the ambient cluster
temperature compared with that expected from simple cooling. This
inconsistency between theory and observations has led to the proposal
of many different mechanisms to reheat the cooling gas.  One popular
subset of these heating models appeals to energy input from the active
nucleus in the central dominant galaxy, which is usually found to
harbour a radio source (\citealt{Burnsetal97}).

The central dominant galaxy in Abell~3581 hosts the powerful radio
source PKS~1404-267. Since this is a cool
cluster with a powerful radio source at its centre, it might be
expected that if active galactic nuclei (AGN) are able to heat the ICM in
clusters, this source might show the effects more clearly than richer
clusters. However, although we will show that the ICM does not have
significant amounts of gas below a temperature of $kT\sim0.8$~keV,
there seems to be no other clear evidence for heating of the ICM by the AGN.

We note that Abell~3581 has a similar mean temperature and luminosity
to the cluster Abell~1983 which was recently studied with \xmm{} by
\citet{PrattArnaud03}. The properties of the inner core of Abell~3581
are however quite different from those of Abell~1983.

\section{Observations}
\label{observations}
We present a \chandra{} observation
(Obsid: 1650, Sequence number: 800118) of the cool cluster Abell~3581
which was made on 2001 June 7 with the S3 back-illuminated detector
in the ACIS-S instrument.

Data processing has been done using the \prog{ciao} package available from the
\chandra{} X-ray Centre while spectral fitting used the \prog{xspec}
package (\citealt{Arnaud96}). The events file was reprocessed to apply the
most appropriate gain file to this observation, which was made with a
focal plane temperature of -120C, and to remove periods of bad background
flaring. This resulted in a total of 7165s of good on-source exposure
from the nominal 7260s observation.
The PI column of the resulting events file was corrected for
the temporal dependence of the gain by using the
\prog{corr\_tgain} package
made available by Alexey
Vikhlinin\footnote{\url{http://cxc.harvard.edu/cont-soft/software/corr_tgain.1.0.html}}.

Analysis of the cluster emission used data processed for best
background discrimination from the `very faint' data mode. Analysis of
the point-source used standard `faint mode' processing since the
point-source counts are moderately affected by pileup.

Throughout this paper, we adopt a redshift for Abell~3581
of $z=0.0218$ \citep{Johnstoneetal98} and a cosmology that assumes a flat universe with
$H_0=70\kmps$ and $\Omega_m$=0.3. We use N Wright's web
page\footnote{\url{http://www.astro.ucla.edu/~wright/CosmoCalc.html}} to
calculate that the angular diameter distance to
this source is 90.9 Mpc and its
luminosity distance is 94.9 Mpc. The linear scale is 441pc per
arcsec. Wherever we quote uncertainties these are at the $1\sigma$
level, unless explicitly stated otherwise. In general we have used the
chi-square statistic to assess goodness of fit and tests for
significance of additional parmeters, but have used the \prog{xspec}
implementation of Cash's C-statistic
to determine parameter uncertainties and confidence
regions\footnote{\url{http://xspec.gsfc.nasa.gov/docs/xanadu/xspec/manual/manual.html}}.

Recently \citet{BarnesNulsen03} have carried out
a detailed investigation of the fluctuations in the Galactic foreground
absorption equivalent Hydrogen column density in the direction of
this cluster. They
reference two absolute measurements
from the literature:  \citet{Starketal92}, using a
2-degree beamsize find $N_{\rm H}=4.5\times 10^{20}\pscm$, while
\citet{HartmannBurton97} using a 35 arcmin beam
found $N_{\rm H}=4.1\times 10^{20}\pscm$. In their investigation of {\it
fluctuations} in the value of $N_{\rm H}$ towards this cluster, using new
data from the Australia Telescope Compact Array,
\citet{BarnesNulsen03}
find that at the
5$\sigma$ level fluctations larger than $\pm6.2\times10^{19}\pscm$ are
ruled out. Throughout this paper we adopt therefore the value of $N_{\rm H}=4.1\times
10^{20}\pscm$ (from the more compact beam observation) as the value of the
Galactic foreground absorption.

\begin{figure}
\protect\resizebox{\columnwidth}{!}
{\includegraphics{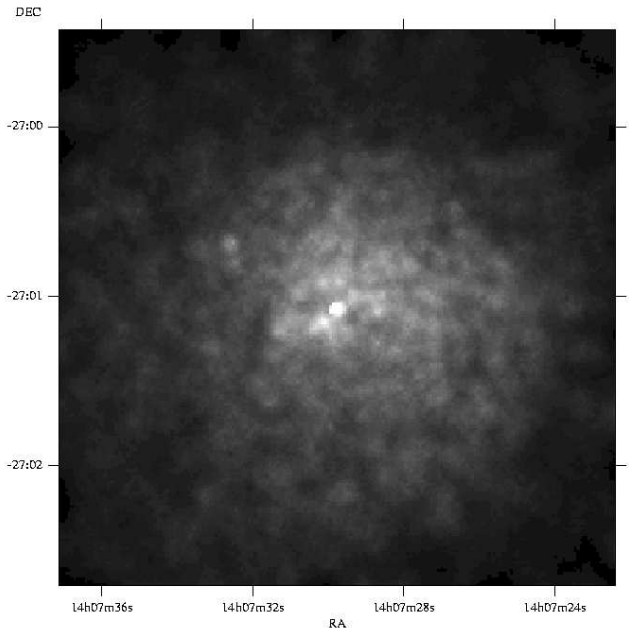}}
\caption{\chandra{} X-ray image of Abell~3581 in the 0.3-3~keV band. The
  data have been binned to 1 arcsec pixels and adaptively smoothed.}
\label{xray}
\end{figure}

In Fig.~\ref{xray} we show the \chandra{} X-ray image of the
central $\sim$3.5x3.5 arcmin region of Abell~3581 in the 0.3-3.0~keV
band. The X-ray events were binned to have 1 arcsec pixels before being
adaptively smoothed. A clear detection of an X-ray emitting active
nucleus is seen close to the centre of the cluster X-ray emission.

\begin{figure}
\protect\resizebox{\columnwidth}{!}
{\includegraphics{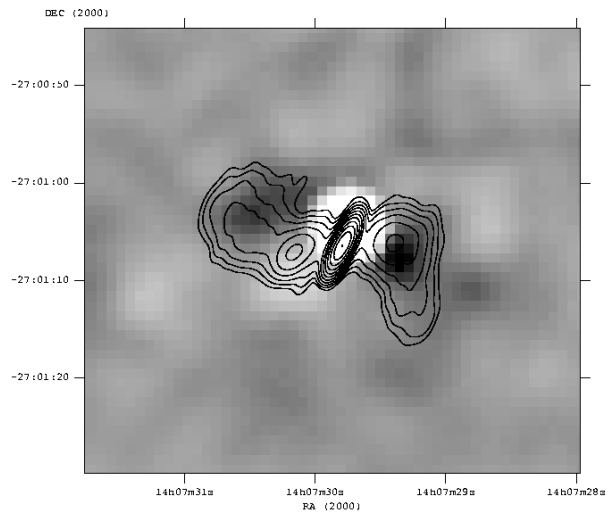}}
\caption{Unsharp masked X-ray image of Abell~3581 (greyscale) with the 20cm VLA map
  of the PKS~1404-267 radio source overlaid. There are 12 contours in
  the radio map spaced equally in logarithmic steps between 0.7~mJy/beam
  and 390~mJy/beam.}
\label{xray-radio}
\end{figure}

The presence of the radio source PKS~1404-267 at the centre of this
cluster has been known since the 1960s, but the \chandra{} image
shows, for the first time, that there is an X-ray emitting point
source associated with it. In Fig.~\ref{xray-radio} we show a close-up
of the X-ray image (which has been unsharp masked)
with a map of the radio source obtained from the
National Radio Astronomy Observatory\footnote{The National Radio
Astronomy Observatory is operated by Associated Universities, Inc.,
under cooperative agreement with the National Science Foundation.}
archive overlaid in contours.  This 700s 20cm (1477~MHz) radio image
was taken on 1987 September 21 with the VLA in the ``A''
configuration. The peak flux density is 396 mJy/beam with a
synthesised beam of 3.34$\times$1.14 arcsec in position angle $-$27
degrees. Clear cavities in the X-ray emission are seen to the East
and West of the nucleus, coincident with the lobes of the radio
source.

\section{The X-ray Point source}

The X-ray point source is located approximately at
RA(2000)=14:07:29.8, Dec(2000)=$-27$:01:04.
We extracted a spectrum from a circular region of
radius 1.8 arcsec centred on the point source. The major component of
the background for this source is the surrounding cluster emission, so
in this case, we extracted the background spectrum from an
annular region surrounding the
nucleus with radius between 1.97 and 3.94 arcsec. Spectra were
rebinned to have a minimum of 20 counts per bin.

Redistribution matrices and ancilliary response functions were
computed using calibration files from the \chandra{} calibration
database version 2.26. This release corrects a previous
one-channel offset error in the FITS
embedded function file and incorporates a correction for
the degradation in
low-energy quantum efficiency due to the build-up of a contaminant on
the filter wheel.

We have modelled the spectrum of the point source as a power-law
modified, initially, only by Galactic absorption. Preliminary fit
results were input to the WebPIMMS%
\footnote{\url{http://heasarc.gsfc.nasa.gov/Tools/w3pimms.html}}
count rate simulator which suggested that
pileup was important at approximately the 8 per cent level. This
is at a level where pileup is starting to become significant, so
all subsequent fitting of the point source was done including
the \prog{xspec} implementation of the pileup model of \citet{Davis01}.

The best fit using the pileup model applied to a power-law with the
Galactic absorption column density fixed at
$4.1\times10^{20}\pscm$ gives a chi-square value of 11.2 for 20
degrees of freedom. Allowing the absorption to be
freely fitted does not significantly decrease the chi-square value.

In order to estimate the power-law index and its
statistical uncertainties we use Cash's C-statistic
(\citealt{Cash79}, implemented in
the \prog{xspec} \texttt{cstat} command) since there are
many PI bins with the minimum number (20) of counts.
The best fitting power-law model has a photon number index of
$2.28^{+0.11}_{-0.15}$, a 2-10~keV flux of $2.16\times
10^{-13}\ergpscmps$ and a
2-10~keV (rest frame) luminosity corrected for Galactic absorption of
$2.35\times 10^{41}\ergps$. The 0.1-10~keV luminosity is $1.08\times
10^{42}\ergps$.

We can calculate the expected luminosity of the point source, assuming
that it is a black hole accreting according to standard Bondi
theory (\citealt{Bondi52})
with a radiative efficiency of 10 per cent. The bolometric luminosity
is given by:
$$\mysub{L}{acc}={{{8.50\times10^{26}}\mysub{M}{bh}\mysub{n}{e}} \over
{T^{1.5}}}\ergps,$$ where \mysub{M}{bh} is the mass of the black hole
in solar masses, \mysub{n}{e} is the electron density in $\pccm$, and
$T$ is the accreting gas temperature in keV.

The work of \citet{Bettonietal03} gives the mass of the black hole in
PKS~1404-267 as being in the range $8.5-9.5\times10^{8}\Msun$ while the
temperature and density of our innermost spatial bin calculated from
the \texttt{projct} model in section 4.2
are $kT=0.81$~keV and $\mysub{n}{e}=0.066 \pccm$ respectively.
These values yield a
bolometric luminosity of $6.9\times10^{43}\ergps$.
If we assume that most of the accretion luminosity comes out in the 0.1-10
keV band then the ratio of observed to expected luminosity is
0.016. The nucleus is underluminous by a factor of $\sim64$. This is
much larger than any expected bolometric correction ($<\times10$). We
note also that part of the `point' source may actually be the
innermost parts of the radio jets, at the nucleus itself.

We note that appropriate values of temperature and
density to put into the formula for the luminosity are those
pertaining at the accretion radius:
 $$\mysub{R}{acc}\sim {G\mysub{M}{bh} \over \mysub{c}{s}^2},$$ where 
$\mysub{c}{s}$ is the sound speed, or
$$\mysub{R}{acc}={3.705\times 10^{-11}\mysub{M}{bh} \over T}.$$
For the innermost
measured value of $T$ in our data $R_{acc}=0.04$~kpc, which is much closer
to the nucleus than our innermost bin can resolve (5~kpc). The work of
\citet{DiMatteoetal03} on M87 has shown that going in from 5~kpc to
the central 1~kpc region the intracluster medium temperature drops by
about a factor of two while the density rises by about a factor of
three. If similar trends occur in PKS~1404-267 they would increase
the expected accretion luminosity, and the discrepancy in the observed
value, by more than a factor of 8.

We note that underluminous accretion is also seen in M87 (in the Virgo
cluster) by \citet{DiMatteoetal03}, NGC~6166 (in Abell~2199) by
\citet{DiMatteoetal01}, and the Sombrero galaxy by
\citet{Pellegrinietal03}.
However, in these objects the luminosity
deficit is rather larger (a factor of $\sim 10^4$) despite the black
hole mass being very similar to the one considered here. Possible
solutions to the discrepancy are discussed by \citet{DiMatteoetal03}.

\section {Cluster Emission}
\begin{figure}
\protect\resizebox{\columnwidth}{!}
{\includegraphics{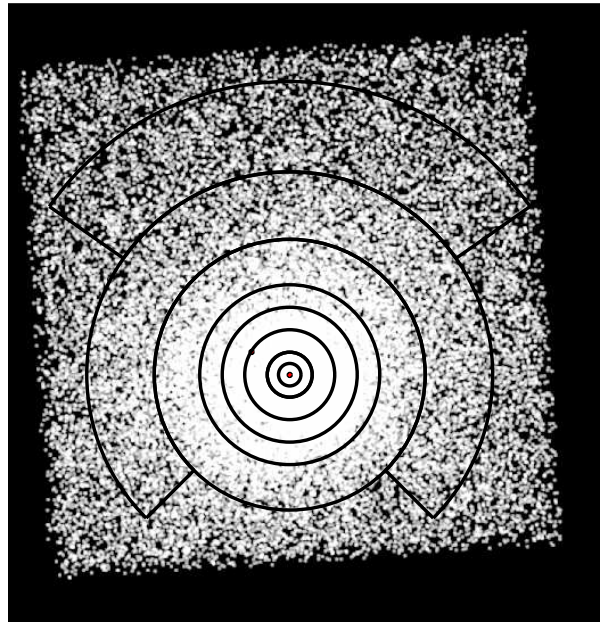}}
\caption{Map of the entire S3 chip showing the concentric
  regions from which spectra were extracted. North is to the top, East
  to the left.}
\label{regions}
\end{figure}

Since the surface brightness distribution of the cluster emission is
quite circularly symmetric we have
extracted spectra from a series of concentric circular annuli centred
on the nuclear point source, but excluding counts from that point
source, and one other at: RA(2000)=14:07:32.6,
Dec(2000)=$-27$:00:41. Fig.~\ref{regions} shows the regions used overlaid on
an image of the cluster. The radio source is contained almost entirely
within the central 5~kpc radius region.

It is not possible to use on-chip data from the same observation as
background since the cluster occupies the whole of the S3 chip. We
therefore reprocessed the standard background field which most nearly
matched the observation date and instrumental configuration to have
the same gain file as the target observation. It was not possible to
apply the time-dependent gain correction to the background events
since the time of observation has been removed from the standard
background fields.

Background spectra were 
extracted from identical chip regions of the
reprocessed background file by reprojecting the detector coordinates
on to the sky coordinates of the target observation.

Redistribution matrices and
ancilliary response files for these spatially extended regions were
made using the \prog{ciao} 3.0.2 tools \prog{mkrmf} and \prog{mkwarf}.
Background spectra
were extracted from the same regions of the standard background fields
as were used to extract the source spectra.

\subsection{Projected quantities}
\begin{figure}
\protect\resizebox{\columnwidth}{!}
{\includegraphics{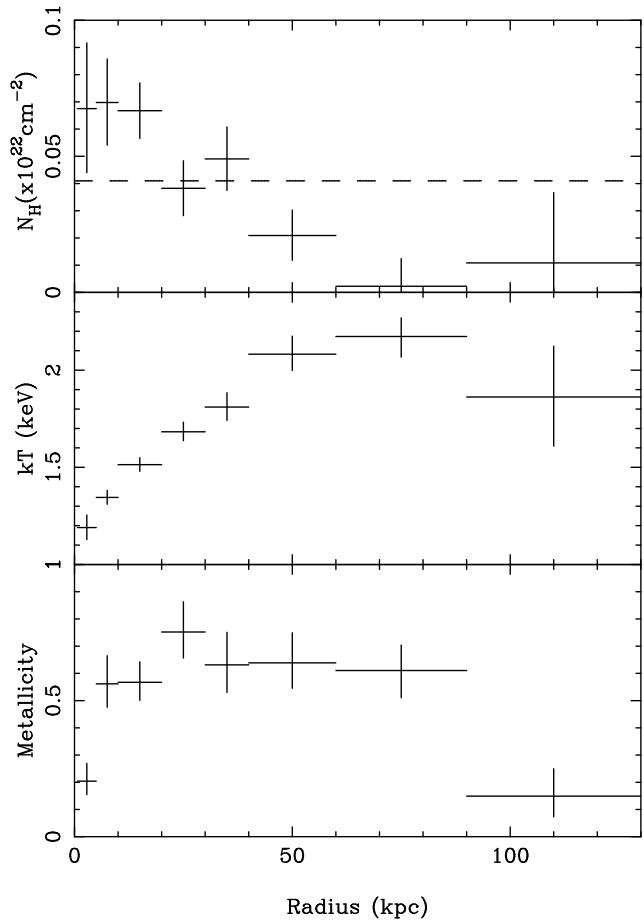}}
\caption{Absorption, temperature and metallicity profiles as seen in
  projection on the sky. The dashed line shows the Galactic column density.}
\label{projprof}
\end{figure}

Single temperature \texttt{mekal} (\citealt{Liedahletal95}) plasma emission models, incorporating freely
fitting Galactic absorption were fitted (separately)
to the spectra extracted from each annulus.
The absorption, temperature, metallicity and normalizations are free
parameters for each annulus.
In Fig.~\ref{projprof} we show the absorption, temperature and metallicity
profiles of the cluster gas derived from these fits. 

The absorption measured in the annuli beyond 40~kpc from the nucleus does
seem to be slightly systematically low compared with the Galactic column
density inferred from 21cm observations which is shown as a dashed
line in Fig.~\ref{projprof}. This may indicate the level of
uncertainty in the model for the correction of the contaminant that
has caused the degradation of the low-energy response in the
\chandra{} data.
We also note that the absorption column density increases roughly
linearly from 75~kpc in towards the centre of the cluster by
$\sim7\times10^{20}\pscm$.

\citet{Johnstoneetal98} found that the central 3~arcmin region from
the ASCA data required a column density of
$9^{+2}_{-1}\times10^{20}\pscm$. Such a large value would be
consistent with inner 40~kpc or 90~arcsec of the \chandra{}
data. However, if the instrumental absorption has been overcorrected
in the current analysis then a greater degree of agreement between the
ASCA and \chandra{} data is to be expected. The work of
\citet{BarnesNulsen03} has shown that the Galactic foreground
21cm emission is constant to within $0.6\times10^{20}\pscm$ ($5\sigma$
upper limit on the fluctuation) in this
region of the sky so it is unlikely that variations in Galactic
absorption can offer a viable explanation for
the increased absorption. The excess absorption seems to be
centred on the cluster which would argue for a cluster origin rather
than a Galactic origin. The nucleus does not however appear to require
any excess absorption (see section 3).

Looking now at the temperature profile, it is clear that the central
region of the cluster is cool, at $kT\sim1.2$~keV and that the
temperature rises by nearly a factor of $\sim2$ to reach a maximum at
75~kpc from the nucleus. In the one region exterior to this the cluster
temperature drops again to near 1.9~keV, although not with high
significance. We note that due to the fact that the cluster is not
centred on the S3 chip the outermost region is only a partial annulus.

The metallicity in the innermost region has a low value of $\sim0.2$
times the
solar value (assuming the solar values of \citealt{Anders89}). In the region
between 5 and 90 kpc from the nucleus we find a much higher value of
$\sim0.6-0.7$ times the solar value. Further out the metallicity
falls quickly to $<0.2$ times the solar value again.

In order to assess whether the fall in temperature beyond 90~kpc is
cluster wide, we have used an accumulative smoothing / contour
binning technique due to J. Sanders (to be described
by Sanders et al. 2004, in preparation) to define spatial bins of
given signal-to-noise ratio for spectral fitting. This allowed us to
drop the assumption of circular symmetry. Briefly, a surface brightness
map was produced from the raw counts using a smoothing
algorithm which accumulated counts around each pixel until a
signal-to-noise
ratio of 15 was reached (``accumulative smoothing"). The binning method
(``contour binning") then grew bins along directions where the surface
brightness is closest to the existing value, until the signal-to-noise ratio
exceeded 30. Finally, bins were constrained to have an edge length less than or
equal to three times the circumference of a circle of the
same area as the bin. Signal-to-noise ratios are calculated taking
into account noise in the background spectrum which is extracted from
identical regions in chip coordinates from the standard background
files.

\begin{figure}
\protect\resizebox{\columnwidth}{!}
{\includegraphics{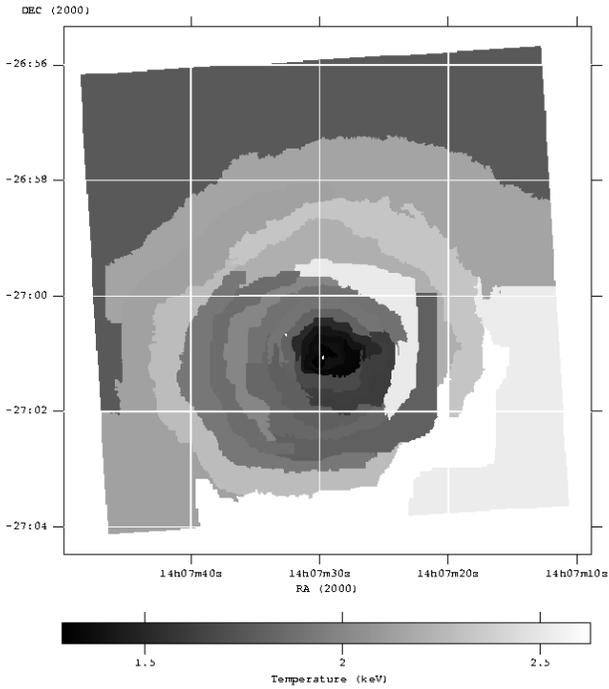}}
\caption{X-ray temperature map of Abell~3581.}
\label{tmap}
\end{figure}

Spectra were extracted from these regions and binned to 20 counts per
spectral bin. New response
matrices and ancilliary response functions were made corresponding to
these regions which were then fit independently using Cash's
C-statistic. The resulting temperature map is shown in Fig.~\ref{tmap}.

It is clear now that the south western part of the cluster is 
systematically hotter than the northern part. Since the outermost
annulus comes only from the northern part of the cluster
it is not surprising that the outermost
annulus has a temperature below the one immediately interior to it since
that partial annulus covers the hotter region.

\begin{figure}
\protect\resizebox{\columnwidth}{!}
{\includegraphics{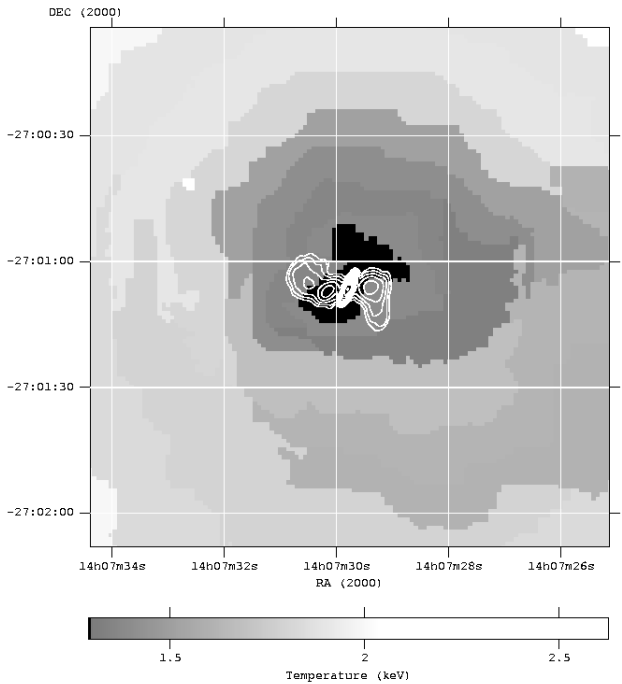}}
\caption{X-ray temperature map of the central region of Abell~3581
  with radio intensity contours overlaid.}
\label{tmapradio}
\end{figure}

In Fig.~\ref{tmapradio} we show a close-up of the temperature map of
the central region of the cluster with the radio intensity contours
overplotted. There is some suggestion that the radio source avoids the
the coolest gas, which may be a direct indication that the radio
source heats the gas. The precise locations of the boundaries of the
individual regions selected for fitting are however somewhat sensitive
to the smoothing and signal-to-noise parameters of the contour accretion
technique, and are required to follow the surface brightness of the
cluster emission.

\subsection {Correcting for projection effects}

We have fit the annular spectra
accounting for projection of counts from the
exterior regions into interior annuli using the \texttt{projct} model in
\prog{xspec}, and assuming spherical symmetry. The volumes of the shells
sampled by each annulus are calculated from FITS header items which
describe the inner and outer boundaries of the annuli. 
We note that the cluster emission extends
beyond the outermost annulus of our analysis so that the 
emitting volume associated with the outermost annulus is too small. This
causes too much projection on to the next annulus in, so that the fit
to that annulus underestimates its brightness.
The resulting effect is minimal beyond the outer two annuli
because the amount of projection is relatively small due to the steep
surface brightness profile; see section 4.3.

The model which we have fitted consists of a single temperature \texttt{mekal}
plasma emission model in which the temperature and metallicity at each
radius are freely fit. A single value of the foreground absorption
column density is also allowed to fit freely and obtains a value
of $3.8\pm0.7\times10^{20}\pscm$, consistent with that expected from
the HI 21cm measurements.

\begin{figure}
\protect\resizebox{\columnwidth}{!}
{\includegraphics{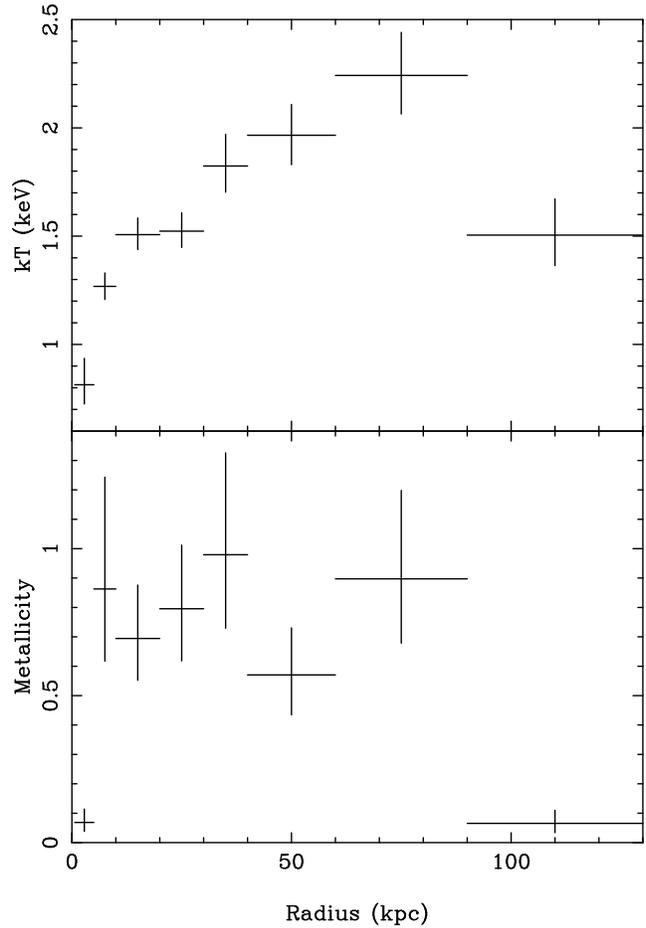}}
\caption{Temperature and metallicity profiles for Abell 3581
derived using the \texttt{projct} model.}
\label{deprojprof}
\end{figure}

In Fig.~\ref{deprojprof} we show the temperature and metallicity
profiles of the cluster gas derived from these fits accounting for projection.
The temperature climbs steadily from 0.8~keV within the central 5~kpc
to around 2.25~keV between 60-90~kpc. We note that the intrinsic temperature of
the central bin is substantially cooler when taking into account the
hotter gas seen in projection against it. In the outermost annulus the
temperature drops back to 1.5~keV. Since this region is only a part of an
annulus, to the north of the cluster (Fig~\ref{regions}), it is not
clear whether this decline in temperature is due to substructure or in
fact a coherent property of the cluster.

The metallicity is consistent with 0.8 solar thoughout most of
the cluster observed in our dataset. The very innermost region
(within 5~kpc, but excluding the nuclear point source)
and the region beyond 90~kpc both have much lower
values of metal abundance, at around 0.1 times the solar value.

The drop in metallicity in the central bin could be due either to an
extended power-law component, or to the iron-bias effect
\citep{BuoteFabian99}. The iron-bias effect results from fitting a plasma
containing more than one distinct temperature component with a single
temperature model.

We have tested for the possibility of there being a spatially extended
power-law or a second \texttt{mekal} component by adding them
(separately) to the baseline \texttt{projct * phabs (mekal)} model
(Model 1), but only in the central bin. The results of these fits are
compared with the baseline model in Table \ref{table1}. For the test
with the power-law component (Model 2), both index and normalization
were left as free parameters while for the test with a second
\texttt{mekal} component (Model 2) we allowed a freely fitting
temperature and normalization but tied the metallicity to that of the
first \texttt{mekal} component. For both tests we found that the
metallicity was very poorly constrained and preferred to fit to
unphysically large values; we have therefore imposed an upper limit of
twice the solar value on this parameter. Indeed the best fit value of
the metallicity hits this limit, but we note that allowing much higher
values of metallicity does not reduce the value of the chi-square
statistic significantly. Our lower limits on the metallicity presented
in \ref{table1} correspond to an increase of 1.0 in the C-statistic
from the value fitted with the metallicity at twice the solar value.

For the model including the power-law component an F-test suggests
that there is a 6 per cent chance of obtaining such a significant drop
in chi-square if the power-law component is \textit{not} present.  For
the model including the second \texttt{mekal} component an F-test
indicates that there is only a 0.7 per cent chance of obtaining such a
significant drop in chi-square if the second \texttt{mekal} component
is \textit{not} present.

\begin{table}
\begin{center}
\caption{
Parameters, for the central spatial bin, of models fitted taking into
account projection effects. Model 1 is \texttt {projct * phabs (mekal)}.
Model 2 is \texttt{projct * phabs (mekal + po)}.
Model 3 is \texttt{projct * phabs (mekal + mekal)}. In models 2 and 3 the power-law
and second \texttt{mekal} components are only present in the innermost
spatial bin. Temperatures are given in keV, abundances are relative to
the solar value and column densities are in units of $10^{20}\pscm$.
``D of F'' is the number of degrees of freedom in the fit.
}
\begin{tabular}{cccc} \\
\hline
\multicolumn{1}{c}{Model} &
\multicolumn{1}{c}{1} &
\multicolumn{1}{c}{2} &
\multicolumn{1}{c}{3}\\
\hline
\mysub{kT}{1}    & $0.81^{+0.13}_{-0.08}$ & $0.64^{+0.07}_{-0.07}$ & $0.61^{+0.06}_{-0.07}$ \\
\mysub{kT}{2}    &  --                    &  --                    & $2.50^{+1.11}_{-0.73}$ \\
Abund            & $0.07^{+0.07}_{-0.03}$ & $>0.14$                & $>0.48$                \\
\mysub{N}{H}     & $4.0^{+0.5}_{-0.5}$    & $3.8^{+0.5}_{-0.5}$    & $3.8^{+0.5}_{-0.5}$    \\
$\chi^2$         & 844.4                 & 837.8                   & 832.7                  \\
D of F           & 714                   & 712                     & 712                    \\
\hline

\end{tabular}
\renewcommand{\baselinestretch}{1.0}
\newline
\label{table1}
\end{center}
\end{table}

We conclude that a two-component \texttt{mekal} model is a
significantly better fit in
the central bin than one with only a single \texttt{mekal} component
and is preferred over one with an extended  power-law component.
However, in both of the two-component models the metallicity is sufficiently
unconstrained that it becomes consistent with that seen in the second to
seventh bins, counting outwards from the centre.

We note that these data are also able to constrain the mass profile of Abell 3581. The
results from this work will be presented as part of a larger sample of cluster
mass profiles by Voigt \& Fabian (2005), in preparation.

\subsection{Testing the projection correction}
\begin{figure}
\protect\resizebox{\columnwidth}{!}
{\includegraphics{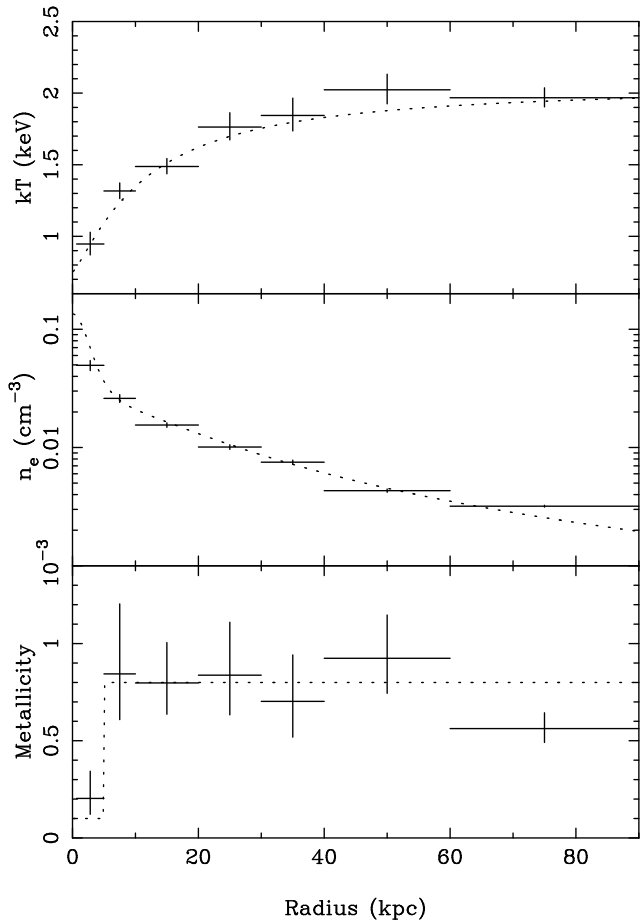}}
\caption{Simulated cluster (dotted lines)
and parameters derived using the \prog{xspec} \texttt{projct} model (crosses).}
\label{simulation}
\end{figure}

We have run a series of tests to check that the \texttt{projct}
\prog{xspec} model
reproduces what is expected when given known synthetic data. We find
that it generally gives good results. In Fig.~\ref{simulation}, the results of
one test which is based upon our dataset are shown.

We start by using the temperature, density and metallicity
profiles appropriate for
Abell~3581, accounting for projection effects, as determined by the
\texttt{projct} model. Since the outermost spatial region shows a significant
temperature drop we have only used data from the central seven spatial
regions in this test. In order to
construct a smooth synthetic cluster
the density profile was fitted with the sum of two beta profiles, while
the temperature profile was fitted with the functional form of
\citet{Allenetal01}. In both cases the outermost point (of the seven)
was omitted
from the fit (see discussion at the end of this section).

We then constructed a spherically symmetric cluster assuming
the ideal gas equation of state. There were 10,000 radial bins equally spaced
out to a radius of 2500~kpc, at which point the cluster was truncated.

Spectra appropriate for the physical conditions at each radius were then
calculated using the \texttt{mekal} plasma code. The emission was integrated
along the line of sight to give a projected spectrum at each radius
in the cluster. Next the spectra were accumulated radially, in the same
annular rings as used in section 4.2, and scaled to the luminosity
distance appropriate to Abell~3581 using our adopted cosmology.
These spectra were then input to \prog{xspec}, multiplied with a
uniform Galactic absorption component, and used to generate
synthetic pulse invariant (PI) spectra assuming the same responses and
integration time as the Abell~3581 observation. The counts were perturbed
following poisson statistics. 

The synthetic spectra were then fitted with the \texttt{projct} model in \prog{xspec}
using the same techniques as in Section 4.2. In Fig.~\ref{simulation}
we have overplotted the synthetic cluster profiles for temperature, density
and metallicity (dotted lines) with those derived from
fitting the PI
spectra with the \texttt{mekal} and \texttt{projct} models, allowing
for Galactic absorption (points with error bars). 
It is clear that the fitted temperature and density
points generally match very well with the `real' synthetic
profiles. There appear to be no systematic residuals in the
temperature or metallicity profiles. In the density profile we note
that the outermost
point derived from the \texttt{projct} fits is very significantly above the
cluster profile. This is because there are counts projected on to this
region that cannot be taken into account correctly. The
\texttt{projct} model requires the projected outer radius for each annulus to
be associated with that data set in order to calculate the volume and
therefore the projection factors. For the outermost region the volume
associated with the counts is wrong in the sense that it is too
low. This leads to an overestimate of the surface brightness and
density in the outermost annulus. Although this causes too much
emission to be subtracted off the next inner annulus, it can be seen
that the effect is very small and the errors due to this effect
quickly damp out. We therefore have confidence in the
projection-corrected
results presented in section 4.2, apart from the density in the
outermost annulus.

\subsection{The issue of entropy}

\begin{figure}
\protect\resizebox{\columnwidth}{!}
{\includegraphics{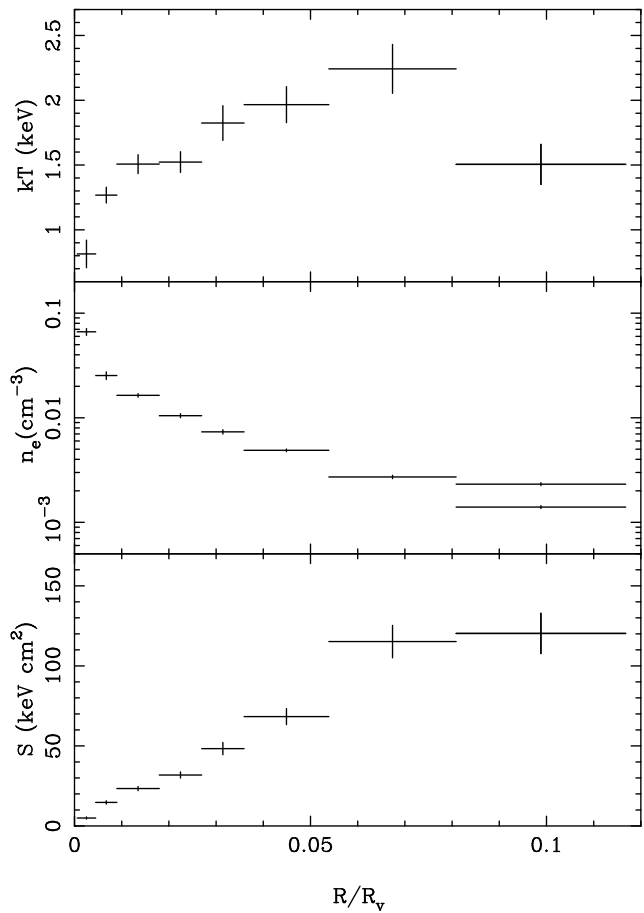}}
\caption{Temperature, electron density, and entropy profiles
  determined from the \texttt{projct} model. The radial coordinate has been
  scaled by the virial radius, 1.11Mpc. There are two points plotted in the
  electron density profile at $0.1\mysub{R}{v}$. The upper point is
  determined directly from the \texttt{projct} value while the lower
  point is determined from an extrapolation of a double-beta profile
  fit to the seven inner points. The entropy point plotted at
  $0.1\mysub{R}{v}$ is calculated using the extrapolated fit value for
  the density. See text for details.
}
\label{entropy}
\end{figure}

In Fig.~\ref{entropy} we show the temperature, electron density and
entropy profiles deduced from the fits accounting for projection
plotted against the scaled radius. Many authors choose to scale the
radial coordinate by $\mysub{R}{200}$ the radius within which the mean
cluster density is 200 times the mean density of the
universe. $\mysub{R}{200}$ has been shown by \citet{ColeLacey96}
to be proportional to the virial
radius, which is the appropriate scaling quantity in hierarchical
self-similar models of structure formation. We have used equation 9 of
\citet{Sandersonetal03} as an approximation to the virial
radius for our scaling. Using $\mysub{T}{cluster}$=2~keV
(a value which is between that
measured in the two outer bins of our data, and within the range
measured by \citealt{Johnstoneetal98}, Table~1) we find that the
virial radius is 1.11Mpc.
\citet{Sandersonetal03} have shown that this equation can
significantly overestimate $\mysub{R}{200}$ particularly in small systems
with large temperature gradients. Since we have used the maximum
temperature in the cluster (at least within the region covered by our
data) rather than an emission-weighted average our $\mysub{R}{200}$ ought to be
reasonably accurate.

In calculating the entropy we have used the definition given 
by \citet{LloydDaviesetal00}, in which the entropy is $$S=T
\mysub{n}{e}^{-2/3}.$$ The outermost spatial bin at $0.1\mysub{R}{v}$
is subject to two
problems. Firstly, as described in section 4.3 the \texttt{projct}
model overestimates the density. This
translates into too small a value of entropy in that bin. In order to
avoid this problem we have estimated the electron density by fitting the other
points in the density profile with a double-beta model and
extrapolated this to the position of the final spatial bin. This
yields an electron density that is a factor of 1.66 times lower than
calculated by \texttt{projct} for that bin. In Fig.~\ref{entropy} the higher
electron density is that determined directly from the \texttt{projct}
model while the lower value is our corrected value.
The entropy point
($S=120\pm13\keVscm$) that we plot at
$0.1\mysub{R}{v}$ is derived from this corrected density.

Secondly, the drop in temperature seen in this bin may not be
representative of the cluster averaged profile since it is only a
partial annulus and there may be azimuthally dependent temperature
structure within the cluster. If the azimuthally averaged temperature
profile were to remain flat at the average of the values in the sixth
and seventh annuli, counting outwards from the cluster centre, it
would have a value of 2.1~keV (however, we note that forcing this
value into the \texttt{projct} model would reduce the temperature of
those inner bins). Values of $kT=1.8-2.2$~keV were found by
\citet{Johnstoneetal98} from fitting various models to the \asca{}
data from the central 3 arcmin radius region ($0.07\mysub{R}{v}$) in
this source. Using the value of 2.1~keV for the temperature would
increase the entropy at $0.1\mysub{R}{v}$ to $167\keVscm$. We consider
this to be an absolute maximum value for this quantity.  We also note
that the virial radius increases as the square root of the cluster
temperature so it is not strongly affected by changes in the adopted
cluster temperature.

The entropy profile for Abell~3581 (as a function of radius scaled by
the virial radius) lies well below the profile for clusters with
similar temperatures as published by \citet{LloydDaviesetal00} (taking
into account the $S\propto h^{-1/3}$ scaling for the different assumed
values of \mysub{H}{0}). This result stands, despite the presence of the
relatively strong Parkes radio source in Abell~3581.  The entropy
profile for Abell~1983 \citep{PrattArnaud03}, which has a similar
temperature ($kT=2.1\keV$), has entropy values which are
$\sim40\keVscm$ higher in the region covered by the \chandra{} data on
Abell~3581 (the entropy scaling factor $(1+z)^{-2} T$ expected from
self-similar formation models, e.g. \citealt{PrattArnaud03},
is the same within 1 per
cent for these two objects). Many authors
(see e.g. references in \citealt{BinneyTabor95}, \citealt{Fabianetal01},
\citealt{Churazovetal01})
have recently argued that the
mechanical heating from an active galactic nucleus with jets can
balance the cooling from high density gas in the cluster and thereby
prevent the formation of a classical cooling flow. Such heating might
be expected to increase the entropy of the clusters, particularly where
the gas temperature is cool and the radio source is strong, but this
is clearly not the case in Abell~3581.

\citet{Ponmanetal03} discuss the entropy of intergalactic gas in
elliptical galaxies, groups and clusters. Their fig.~4 shows the gas
entropy at $0.1 \mysub{R}{200}$ as a function of temperature for 66
virialized systems. Below about 4~keV there is a wide spread in this
fiducial entropy. The value of entropy at $0.1\mysub{R}{v}$ in Abell~3581 
calculated from the data at that
position in the cluster (with the corrected value for the electron
density) is $120\pm13\keVscm$, or $167\pm18\keVscm$, if we assume the
high value of the temperature in the outermost annulus.
By either of these measures,
Abell~3581 has an entropy at $0.1\mysub{R}{v}$ significantly below that for
the mean of clusters at 2~keV $\sim233\pm30\keVscm$, as shown in fig.~5 of
\citet{Ponmanetal03}. Fig.~4 in that same work shows there is a large
range in entropy for clusters at 2~keV. Our measured value for
Abell~3581 would make it equal to their lowest entropy cluster
(Abell~262) at 2~keV. Abell~3581 emphasizes the overall spread in properties of
low temperature clusters.

A power-law fit to the entropy as a function of radius
($S=kr^{\alpha}$) gives a slope $\alpha=0.87\pm0.03$. Further
including a constant offset in the entropy at zero radius does not
produce a significant reduction in the chi-square statistic as
determined using an F-test. We note that this is a much flatter slope
than expected for pure shock heating of the intracluster gas which is
expected to give a slope of $\alpha=1.1$
\citep{TozziNorman01}. However, \citet{Ponmanetal03} have shown, in
their fig. 3, that such slopes are usually only seen beyond
$0.2\mysub{R}{v}$, which is outside the region covered by our
data. The flattening of this slope in the inner regions of clusters
may be due to the cooling out of the lower entropy gas.

\subsection {A classical cooling flow?}
We extracted a spectrum from the inner 3~arcmin (or 79~kpc) of the
cluster (but excluding the point source) for
comparison with the ASCA spectrum of
\citet{Johnstoneetal98}.
This region is expected to contain most of the mass deposition if
a classical cooling flow is present \citep{Johnstoneetal98}.
We have fitted a range of models to this region; the results are presented in Table \ref{table2}.
Initially we fitted a single temperature \texttt{mekal} model affected only by (freely
fitting) Galactic absorption (Model 1). This is our baseline model for the region.

\begin{table}
\begin{center}
\caption{
Parameters of fits to cooling flow region. Model 1 is \texttt{phabs (mekal)}.
Model 2 is \texttt{phabs (mekal + mekal)}.
Model 3 is \texttt{phabs (mekal + mkcflow)} with the lower temperature fixed at 0.08~keV.
Model 4 is \texttt{phabs (mekal + mkcflow)} with the lower temperature
allowed to fit freely. Temperatures are given in keV, abundances are
relative to the solar value, the mass deposition rate ${\dot M}$ is given
in solar masses per year and column densities are in units of $10^{20}\pscm$.
``D of F'' is the number of degrees of freedom in the fit.
}
\begin{tabular}{ccccc} \\
\hline
\multicolumn{1}{c}{Model} &
\multicolumn{1}{c}{1} &
\multicolumn{1}{c}{2} &
\multicolumn{1}{c}{3} &
\multicolumn{1}{c}{4}\\
\hline
\mysub{kT}{upper}             & $1.68^{+0.02}_{-0.02}$ & $1.86^{+0.05}_{-0.05}$ & $1.83^{+0.04}_{-0.04}$ & $1.90^{+0.06}_{-0.05}$ \\
\mysub{kT}{lower}             &  --                    & $0.86^{+0.09}_{-0.07}$ & 0.08                   & $0.60^{+0.11}_{-0.11}$ \\
Abund                         & $0.53^{+0.03}_{-0.03}$ & $0.75^{+0.06}_{-0.06}$ & $0.74^{+0.07}_{-0.06}$ & $0.76^{+0.07}_{-0.06}$ \\
${\dot M}$                    &  --                    &  --                    & $8.1^{+1.3}_{-1.3}$    & $12.7^{+3.9}_{-2.8}$ \\
\mysub{N}{H}                  & $4.4^{+0.4}_{-0.4}$    & $3.9^{+0.5}_{-0.5}$    & $5.1^{+0.5}_{-0.5}$    & $4.0^{+0.5}_{-0.5}$    \\
$\chi^2$                      & 338.2                  & 303.9                  & 306.2                  & 302.8                  \\
D of F                        & 205                    & 203                    & 204                    & 203                    \\
\hline

\end{tabular}
\renewcommand{\baselinestretch}{1.0}
\newline
\label{table2}
\end{center}
\end{table}

Adding in a second \texttt{mekal} component (Model 2) in which the metallicity is linked
to the first component reduces the chi-square value to 303.9 for 203
degrees of freedom and gives a significantly better fit (F-test
significance=$1-1.9\times10^{-5}$).

Next we have substituted a cooling flow model for the second
\texttt{mekal} component (Model 3) in
which the upper temperature and metallicity of the cooling flow are
tied to that of the \texttt{mekal} component, and the lowest
temperature in the cooling flow model is
below the \chandra{} band. Although this model is not a better fit than the
two-temperature model in terms of the chi-square statistic, it only introduces
one further fit parameter instead of two for Model 2 and an F-test shows
the rediction in chi-square for this model to have a greater level of
significance ($1-6.4\times10^{-6}$).
The mass deposition rate fits to 
$8.1\pm1.3\Msunpyr$. (Note that for comparison with previous values of
mass deposition rate published by \citet{Johnstoneetal98} this
value should be doubled to account for the different assumed value
of $H_0$.)

\begin{figure}
\includegraphics[angle=270,width=\columnwidth]{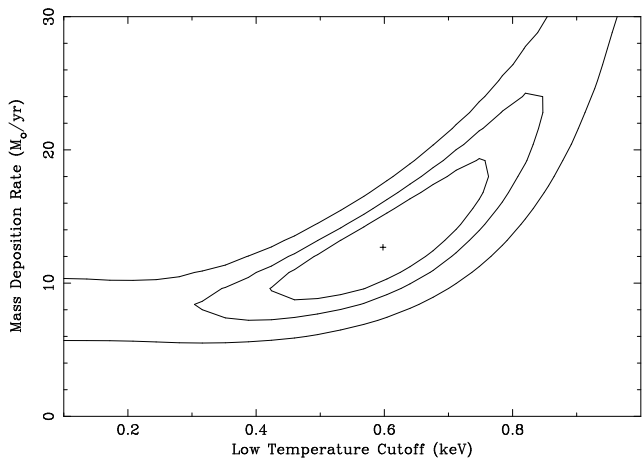}
\caption{Allowed values of mass deposition rate as a function
  of low temperature cutoff in the cooling gas. Contours are plotted
  at $\Delta$ C-statistic = 2.3, 4.61 and 9.21 which correspond to 68, 90 and
  99 per cent confidence regions for two interesting parameters.}
\label{mdotplot1}
\end{figure}

Finally, allowing the lower temperature of the cooling flow
to be a free parameter allows chi-square to drop further to 302.8. There is an
F-test probability of 0.15 of obtaining such a large drop
in chi-square if the additional fit parameter were not required. The
allowed mass deposition rate is now $12.7^{+3.9}_{-2.8}\Msunpyr$.
In Fig.~\ref{mdotplot1} we show
contours of constant C-statistic in the mass depostion rate / lower
temperature plane. The three contours have $\Delta$
C-statistic = 2.3, 4.61 and 9.21 which correspond to 68, 90 and 99 per cent
confidence regions for two interesting parameters.

\citet{Johnstoneetal98} performed a surface brightness deprojection
of the {\it ROSAT} HRI data. Within the central 3~arcmin (110~kpc in
$H_0=50$, $q_0=0$ used in that paper), a mass deposition rate of
$45-50\Msunpyr$ was found. Scaling these values to our current
cosmology would give a value of $22-25\Msunpyr$, around a factor of
three higher than fitted in this section when the gas is allowed to
cool to very low temperatures. Similar results have been
found for many other clusters by e.g.
\citet{Petersonetal01}, \citet{Petersonetal03}, \citet{Kaastraetal04}.
a lower cutoff temperature to the
cooling gas proposed.
At a cutoff temperature of around $\sim 0.8$~keV agreement
with the deprojection mass deposition rates is achieved, albeit with a
poor chi-square value. We note that 0.8~keV is the lowest temperature
directly observed in the \texttt{projct} fits of section 4.2.

The problem remains as to why the gas appears to cool by about a
factor of $\sim 2.5$ yet not any further and does not accumulate at
the lower temperature.

\section{Discussion}
So far we have considered heating in Abell~3581 only from the point of
view of balancing the cooling from the ICM in the core of the cluster.
However, there is a more global property of clusters that shows that
heating has been important. The observed luminosity/temperature
($L_x/T_x$) relation for clusters, where $\mysub{L}{x}\propto \mysub{T}{x}^3$
(e.g. \citealt{ArnaudEvrard99}) rather than $\mysub{T}{x}^2$
as expected if only
gravitational heating is involved (\citealt{Kaiser86}), indicates that
some form of extra heating has taken place. The level corresponds to
about 2-3~keV per particle (e.g. \citealt{Wuetal00}).  Low temperature
clusters and groups show a large dispersion in the
$\mysub{L}{x}/\mysub{T}{x}$ relation
(\citealt{HelsdonPonman00}), which may mean that although heating is
widespread, it is not the same in all objects. One possible origin for
the heating, which has been suggested by many authors, is the radio
source which is often associated with the central galaxy.

We have measured the unabsorbed X-ray luminosity in the central 60~kpc
of both the Perseus cluster and Abell~3581 (excluding the point
sources) and find values of $1.5\times10^{44}\ergps$ and
$1.4\times10^{43}\ergps$ respectively in the 0.3-10~keV band, so the
Perseus cluster is factor of ten brighter than PKS~1404-267 in this region.
By comparison, the 1.4~GHz radio luminosity of the inner two radio
lobes in the Perseus cluster is $2.0\times10^{24}\WHz^{-1}$ while the
1.4~GHz radio luminosity of the lobes in PKS~1404-267 is
$2.5\times10^{23}\WHz^{-1}$, showing that the distributed radio power in the
Perseus cluster is also a factor of ten brighter than in
PKS~1404-267.

Abell~3581 is therefore particularly interesting as it is a low
temperature cluster with a strong radio source where heating seems not
to have been effective, even close to the active nucleus.
\citet{Roychowdhuryetal04} have suggested
that heating by the AGN should also affect the entropy profile of a
cluster beyond the coolest parts. Even if the 
radio source has offset radiative cooling in the very innermost
regions, it is clear that it has not had a strong influence
on the gas entropy at $0.1\mysub{R}{200}$, compared with similar low
temperature clusters.

Various explanations for this are possible. a) The radio source may be
very young compared with the radiative cooling
time of hundreds of Myrs, in which case radiative cooling would have
dominated in the past. However,
no rapidly cooling cluster has yet been found
elsewhere. b) The major
heating phase in clusters may have occurred 5 or more billion years ago,
beyond $z\sim1$, when the central active galaxy or galaxies were much more
important. This may have been a variable process which has left some clusters
without significant heating. c) Perhaps Abell~3581 is particularly
old and thus dense, and has had few recent mergers. Cooling in its high
density core may absorb all the kinetic energy produced by the present
active nucleus so that little is available for the rest of the
gas. d) The dense core may have been introduced into the
cluster as a result of a merger, or e) The dense cool core may have caused the
moderately powerful radio source by more effectively fuelling the
central black hole. Radio source heating as inferred for the Perseus
cluster (by the
dissipation of sound waves; \citealt{Fabianetal03}) may also
apply, and scale, to Abell~3581.

The core of Abell~3581 appears to be similar to that of the Perseus cluster,
except that it is about an order of magnitude less luminous and has a
temperature that is about one third of that in the Perseus cluster.
Somehow the core has avoided the heating,
or at least the effects of heating, common to most other low temperature
clusters and groups.

\section*{Acknowledgments}
ACF acknowledges support by the Royal Society. We thank J. Sanders for
the use of his accumulative smoothing / contour binning software and
many helpful discussions. We also thank an anonymous referee for
helpful comments.

\bsp

\label{lastpage}

\end{document}